\documentclass[%
 reprint,
superscriptaddress,
showpacs,
showkeys,
 amsmath,amssymb,
 aps,
pra,
floatfix,
]{revtex4-1}

\usepackage{graphicx}
\usepackage{dcolumn}
\usepackage{bm}
\usepackage{amsmath}
\usepackage{mathtools,calc}
\usepackage{lipsum}
\usepackage{braket}
\usepackage{amssymb}
\usepackage{MnSymbol}
\usepackage{csquotes}
\usepackage{empheq}

\usepackage{color}
\usepackage{changes}
\usepackage{hyperref}
\usepackage{cleveref}
\usepackage{units}
\usepackage{comment}
\begin{document}

\preprint{APS/123-QED}

\title{Decomposition of the transition phase in multi-sideband RABBITT schemes }

\author{Divya Bharti}
\affiliation{Max-Planck-Institute for Nuclear Physics, D-69117 Heidelberg}
\author{David Atri-Schuller} 
\author{Gavin Menning}
\author{Kathryn R.~Hamilton}
\affiliation{Department of Physics and Astronomy, Drake University, Des Moines, IA 50311, USA}
\author{Robert Moshammer}
\affiliation{Max-Planck-Institute for Nuclear Physics, D-69117 Heidelberg}
\author{Thomas Pfeifer}
\affiliation{Max-Planck-Institute for Nuclear Physics, D-69117 Heidelberg}
\author{Nicolas Douguet}
\affiliation{Department of Physics, Kennesaw State University, Marietta, GA 30060, USA}
\author{Klaus Bartschat}
\affiliation{Department of Physics and Astronomy, Drake University, Des Moines, IA 50311, USA}
\author{Anne Harth}
\altaffiliation[ ]{anne.harth@mpi-hd.mpg.de}
\affiliation{Max-Planck-Institute for Nuclear Physics, D-69117 Heidelberg}
\date{\today}

\begin{abstract}
Reconstruction of Attosecond Beating By Interference of Two-photon Transitions (\hbox{RABBITT}) is a technique that can be used to determine the phases of atomic transition elements in photo\-ionization processes.
In the traditional \hbox{RABBITT} scheme, the so-called ``asymptotic approximation'' considers the measured phase as a sum of the
Wigner phase linked to a single-photon ionization process and the continuum-continuum (cc) phase associated with further single-photon transitions in the continuum.
In this paper, we extend the asymptotic approximation to multi-sideband \hbox{RABBITT} schemes. 
The predictions from this approximation are then compared with results obtained by an {\it ab initio} calculation based on solving the time-dependent Schr\"odinger equation for atomic hydrogen.
\end{abstract}
\maketitle
\section{Introduction}\label{sec:Intro}
The \hbox{RABBITT} technique was originally introduced for the temporal characterization of atto\-second pulse trains (APTs) 
produced via High-Harmonic Generation (HHG)~\cite{Paul1689}. The utility of this
technique was later extended to measure the relative photo\-ionization time delay from different valence shells in Argon~\cite{PhysRevLett.106.143002}. 
Nowadays, \hbox{RABBITT} is extensively employed to study the attosecond dynamics in atoms 
\cite{Isinger893,PhysRevLett.106.143002, guenot2012photoemission, RABBIT_atomic_comp}, molecules~\cite{RABBITT_CO2, RABBITT_N2}, 
and solids~\cite{RABBITT_gold_silver, RABBITT_copper_expt, RABBITT_copper_comp}. \par 
RABBITT is a pump-probe inter\-ferometric technique, in which an extreme ultraviolet (XUV) APT ionizes 
a target gas in the presence of a time-delayed near-infrared (NIR) pulse, and the kinetic-energy spectra of 
the photo\-electrons are recorded as a function of the varied relative delay. 
Without the NIR probe photon, the photo\-electron spectrum consists of discrete peaks (here also termed  ``harmonics'') corresponding to the high-order harmonic peaks in the XUV spectrum. 
The presence of the probe field leads to the appearance of sidebands as additional peaks between the discrete harmonic peaks. 
In the traditional \hbox{RABBITT} scheme, only one sideband is formed between two consecutive harmonic peaks.  
This requires a minimum of one bound-continuum (bc) transition and one continuum-continuum (cc) transition.
For RABBITT, two paths leading to the same sideband interfere.
As the time delay ($\tau$) between the pump and probe pulse is varied, the sideband signal is periodically modulated.
The phase of this delay-dependent modulation is written as
\begin{equation}
\Delta\phi= \Delta\phi_{XUV} + \Delta\phi_{atom},
\label{eq:0-general}
\end{equation}
where $\Delta\phi_{XUV}$ is the phase related to the average group delay of the XUV pulses. 
In many cases, the atomic phase $\Delta\phi_{atom}$ can be conveniently split into two contributions:
\begin{equation}
\Delta\phi_{atom}= \Delta\eta+\Delta\phi^{cc}.
\label{eq:1-SB_AsymApprx}
\end{equation}
The Wigner-like phase shift $\Delta\eta$ originates from XUV-driven bc-transition processes, 
while the cc-transition phase $\Delta\phi^{cc}$ is associated with additional absorption and emission of a probe photon by the photo\-electron. 
This relation is well accepted and was derived by Dahlstr\"om {\it et al.}~\cite{DAHLSTROM2013} using the ``asymptotic approximation'',
in which the asymptotic form of the scattering wavefunction is used to calculate the two-photon ionization amplitude.
In this approximation, $\phi^{cc}$ becomes universal, as it depends on  
the charge of the residual ion, the photoelectron's kinetic energy, and the probe frequency, but not on the details of the atomic system. 
Dahlstr\"om {\it et al.}~\cite{DAHLSTROM2013} formulated 
an analytical expression of $\phi^{cc}$, which is additionally independent of the angular momentum~$\ell$. 

However, the actual $\phi^{cc}$ does depend on the angular momenta~\cite{DAHLSTROM2013,ivanov2016angular,Fuchs:s}, and this dependence becomes significant close to the ionization threshold.
Since the $\ell$-dependence of $\phi^{cc}$ decreases with increasing photo\-electron energy,  we will generally neglect it in the discussion below, except for
pointing out occasions where this dependence may become important.

For the one-sideband \hbox{(1-SB)} \hbox{RABBITT} setup, the probe-field intensity is usually kept low to avoid significant contributions from multiple cc-transitions.
Increasing this intensity leads to the formation of higher-order sidebands, which then may overlap with the harmonic bands~\cite{RABBITT_intensity_dependence}.
Lately, other forms of \hbox{RABBITT} schemes comprising more than one sideband between two consequent harmonic photoelectron peaks were also proposed and realized.  
For example, a \hbox{2-sideband} \hbox{(2-SB)} \hbox{RABBITT} configuration was used in an attosecond-pulse shaping measurement~\cite{Maroju2020}, 
and an experimental technique for a \hbox{3-sideband} \hbox{(3-SB)} \hbox{RABBITT} scheme was proposed to extract the cc-related phase {\it separately\/} by 
cancelling contributions from the Wigner phase~\cite{Harth}.

In all these other schemes, multiple cc-transitions are involved.  
However, a similar descriptions and interpretation like Eq.~(\ref{eq:1-SB_AsymApprx}) for multi-photon continuum transitions is by no means obvious. 
In this paper, we introduce a decomposition approximation by extending the asymptotic approximation to higher-order matrix elements as mentioned by Dahlstr\"om {\it et al.} in \cite{Dahlstr_m_2012}.

This decomposition approximation leads to the interpretation that the final RABBITT phase is built up from the phases of stepwise 
transitions of the photo\-electron, i.e., first the XUV-induced bc-transition, and then subsequent cc-transitions, each involving a single IR photon.

This manuscript is structured as follows.  In Sect.~\ref{sec:General}, we outline the basic equations on which the paper is built. 
This is followed by Sect.~\ref{sec:Cases}, where we first discuss the well-known \hbox{1-SB} case, before we 
introduce the decomposition approximation, which is then formally applied to the \hbox{3-SB} case.
Details of our derivation are given in the Appendix. Section~\ref{sec:Numerics} provides a brief summary of the {\it ab initio} numerical calculations, against which the predictions
of the decomposition approximation are tested and discussed in Sect.~\ref{sec:Results}.
We finish with a summary and an outlook regarding potential consequences for future experiments.

\section{General formulation}\label{sec:General}

The \hbox{RABBITT} technique is based on the interference of different quantum paths leading to the same final energy. 
All equations in the present paper are written in the non\-relativistic single-active electron (SAE) picture, assuming that
the initial bound electron has orbital angular momentum~$\ell_i = 0$ (often omitted in the notation below), and 
considering linearly polarized electric fields. Unless indicated otherwise, atomic units are used.

In the framework of time-dependent perturbation theory, the general expression of the transition
amplitude describing a quantum path from an initial state $\left|i \right\rangle$ to a final state $\left|\vec{k}_N\right\rangle $ with asymptotic momentum $\vec{k}_N$,
upon absorption of one XUV-pump photon ($\Omega$) and ($N\!-\!1$) IR-probe photons ($\omega$), is given by
the $N$th-order matrix element~\cite{Faisal1987}
\label{eq:amplitude2}
\begin{equation}\label{eq:G+}
M_P^{(N)}\!(\vec{k}_N)\!=\!-i\tilde{E}_{\Omega}\tilde{E}^{N-1}_{\omega} \langle \vec{k}_N|z\!\!\prod_{n=0}^{N-2}[G^+\!(\epsilon_{i}+\!\Omega\!+n\omega)z]|i\rangle,
\end{equation}
where $G^+$ is the retarded resolvent of the free-field hamiltonian, $\epsilon_{i}$ is the initial state energy, $z$ is the electric dipole operator, and $P$ specifies the
path by which the final state is reached.  
 
By projecting the final continuum states for the photoelectron on a partial-wave basis, the matrix elements corresponding to the different angular-momentum channels can be disentangled as 
\begin{equation}
M_P^{(N)}(\vec{k}_N)=  \tilde{E}_{\Omega}\tilde{E}^{N-1}_{\omega} \sum_{\ell_N} {\cal M}^{(N)}_{P,\ell_N}(k_N) Y_{\ell_N,0}(\hat{k}_N). \label{eq:amplitude_2}
\end{equation}
The sum over $\ell_N$ represents the possible orbital angular momenta in the final state, and $Y_{\ell_N,0}$ are spherical harmonics. 
Furthermore, \hbox{$\tilde{E}_{\Omega}=E_{\Omega}{\rm e}^{{i}\, \phi_\Omega}$} and \hbox{$\tilde{E}_\omega=E_\omega{\rm e}^{{i}\, \omega \tau}$}  (for absorption)
are the complex electric-field amplitudes of the  XUV-pump ($\Omega$) and NIR-probe ($\omega$) pulses, respectively. 

Although much effort has been put into estimating multi-photon transition matrix elements \cite{LAMBROPOULOS197687,PhysRevA.73.043408},  
it remains challenging to accurately calculate the phases of multi-photon transition elements for a general target other than atomic hydrogen,  even for $N=2$.
Hence, finding a suitable approximation seems highly desirable.

\begin{figure*}
    \centering
	\includegraphics[width=18.0cm,keepaspectratio]{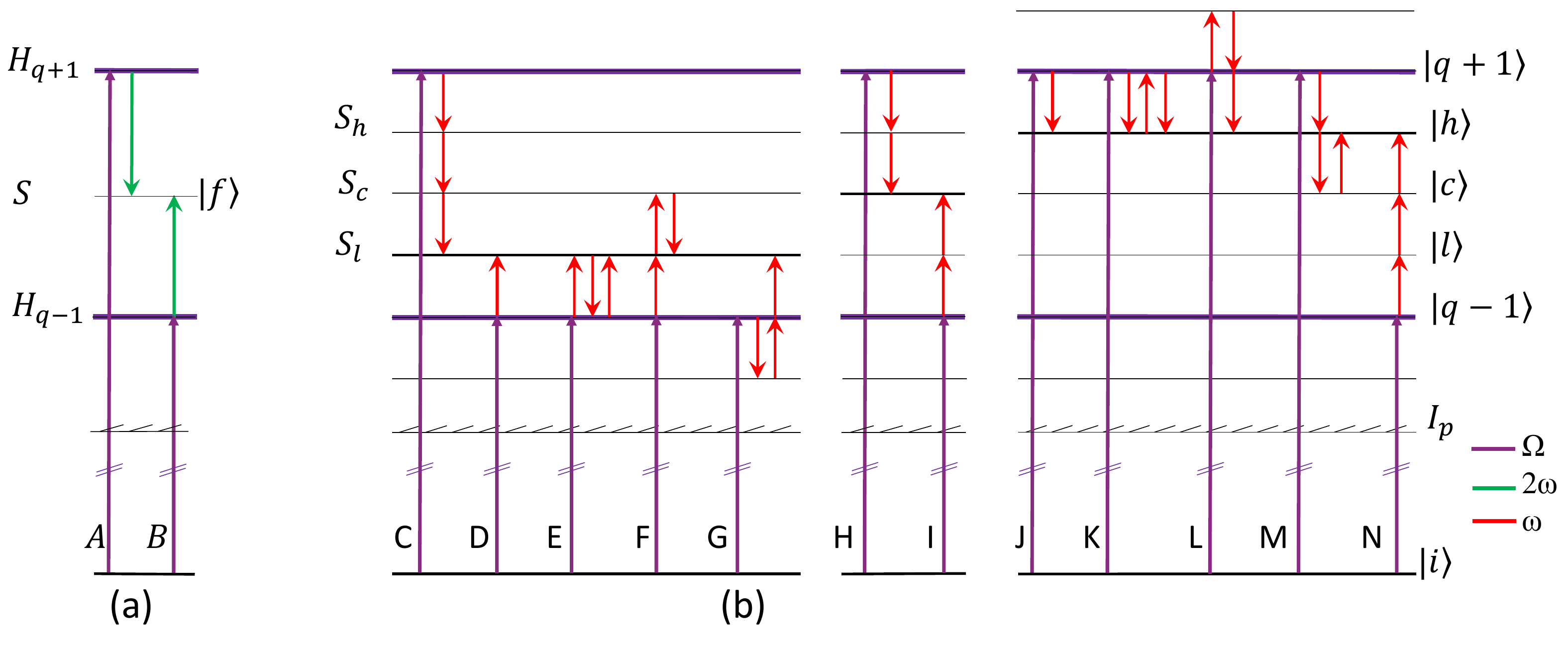}
	\caption{Energy level schemes in (a) traditional \hbox{RABBITT} and (b)  \hbox{3-SB} \hbox{RABBITT}. 
	Photo\-ionization by an XUV-APT results in the appearance of harmonic peaks $H_{q-1}$ and $H_{q+1}$ (with $q$ being an even integer) 
	in the photo\-electron spectrum. The XUV spectrum $(\Omega)$ contains only odd harmonics of the $2\,\omega$ pulse.
	Exchange of NIR probe photons of frequency $2\, \omega$ or $\omega$ leads to the development of (a)~a single sideband or (b)~three sidebands, respectively. 
	Interference among the multiple quantum paths leading to final states with the same energy gives 
	rise to periodic oscillations in the sideband signal as a function of delay between pump and probe. 
	Only the lowest-order paths required to explain the oscillations in the signal are shown. 
	Paths in which the probe photon is absorbed before the pump photon are ignored.}	
	\label{fig:trans}
\end{figure*}

\section{Decomposition of the RABBITT phase}\label{sec:Cases}
We start this section by applying the asymptotic approximation to a hydrogenic system to simplify the 2nd-order matrix element $M^{(N=2)}$. 
We then extend the ideas behind the asymptotic approximation to higher-order matrix elements to arrive at a decomposition relation (see the Appendix for details) and subsequently 
apply it to the \hbox{3-SB} \hbox{RABBITT} case.

\subsection{1-SB \hbox{RABBITT}}\label{subsec:1SB}
Figure~\ref{fig:trans}~(a) shows the energy level diagram and transition pathways involved in a traditional (second-harmonic) \hbox{1-SB} \hbox{RABBITT} scheme. 
To explain the appearance of the sideband~$S$, one needs to consider only two paths, 
A and B, which can both be described by 2nd-order electric dipole transitions. 
Path~A corresponds to the absorption of one XUV photon of energy \hbox{$\Omega_{q+1}=(q+1)\,2\,\omega$} from the harmonic $H_{q+1}$ and emission of one probe ($2\,\omega$) photon, 
while path~B corresponds to the absorption of an XUV photon of energy \hbox{$\Omega_{q-1}=(q-1)\,2\,\omega$}  from the lower harmonic $H_{q-1}$ and the absorption of a probe photon.
Both paths lead to the same final continuum state $\ket{f}$ with asymptotic photoelectron momentum $k_f$, thus resulting in the sideband~$S$.
Using the notation of Eq.~(\ref{eq:amplitude_2}), the two-photon transition amplitudes for paths A and B are expressed as
\begin{eqnarray}
M^{(2,e)}_A(\vec{k}_f)&= &\tilde{E}^*_{2\omega} \tilde{E}_{q+1} \sum_{\ell} {\cal M}^{(2,e)}_{A,\ell}(k_f) Y_{\ell}(\hat{k}_f); \\
M^{(2,a)}_B(\vec{k}_f)&= &\tilde{E}_{2\omega} \tilde{E}_{q-1} \sum_{\ell} {\cal M}^{(2,a)}_{B,\ell}(k_f) Y_{\ell}(\hat{k}_f).
\end{eqnarray}
The superscript $e$ ($a$) indicates the pathway where a probe photon $(2\omega)$ is emitted (absorbed). $\ell (=\ell_f)$ is the angular momentum of the final state $(\ket{f})$. 

The angle-resolved sideband signal is formed as the coherent sum of all the quantum paths leading to the same final momentum state:
\begin{align}
S(\tau,\vec{k}_f) &\propto  \left|M_{A}^{(2,e)}(\vec{k}_f)+M_{B}^{(2,a)}(\vec{k}_f)\right|^{2}. \label{eq:SB_angle_dep}
\end{align}
The phase difference between the absorption and emission paths is varied by changing the time delay ($\tau$) between the pump and the probe pulses. 
This results in an oscillation of the sideband signal as a function of the relative pulse delay.
In the following, we discuss only angle-integrated cases, which results in Eq.~(\ref{eq:SB_angle_dep}) becoming an incoherent sum of partial-wave contributions, i.e.,
\begin{align}
S(\!\tau,k_f\!)\! &\propto\! \sum_\ell \left|\tilde{E}_{q+1}\tilde{E}^*_{2\omega} {\cal M}^{(2,e)}_{A,\ell}(k_f)\! +\! \tilde{E}_{q-1}\tilde{E}_{2\omega} {\cal M}^{(2,a)}_{B,\ell}(k_f) \right|^{2} \nonumber \\
& \propto I_0 + I_1 \, \cos(4\,\omega\tau -\Delta\phi_{XUV}-\Delta\phi_{atom}),
\end{align}
where $\Delta\phi_{XUV} = (\phi_{q+1} -\phi_{q-1})$ is the phase difference between the  harmonic fields $(q+1)$ and~$(q-1)$, while
\begin{equation}
\Delta\phi_{atom}=\mbox{arg}\left[\sum_{\ell}{\cal M}^{(2,e)}_{A,\ell}{\cal M}^{*(2,a)}_{B,\ell}\right]
\end{equation} 
is the atomic phase.

The atomic phase contains the combined effect of the ionizing XUV pump and the NIR probe. 
It is not obvious whether the two contributions can be separated to recover the pure (Wigner only) photo\-ionization delay. 
For the \hbox{1-SB} \hbox{RABBITT} scheme, Dahlstr\"om {\it et al.}~\cite{DAHLSTROM2013} showed that the atomic phase can be split into 
a single-photon ionization phase ($\eta$) and the measurement-induced cc-transition phase ($\phi^{cc}$) when using the asymptotic approximation. 
They also derived an analytical expression for $\phi^{cc}$ corresponding to the single-photon free-free transition of the photo\-electron in the vicinity of the parent ion:
\begin{equation}\label{eq:phi_cc}
\phi^{cc}_{k,\kappa}=\mbox{arg}\left[ \frac{(2\kappa)^{{i}Z/\kappa}}{(2k)^{{i}Z/k}}\frac{\Gamma[2+{i}Z(1/\kappa -1/k)]+\gamma(k,\kappa)}{(\kappa -k)^{{i}Z(1/\kappa -1/k)}} \right].
\end{equation}
\noindent Here $\kappa$ is the wave number of the initial state in the continuum 
while $k$ 
is the wave number of the final photo\-electron momentum after the exchange of an NIR photon, $Z$ is the remaining charge on the parent ion, 
and $\gamma$ denotes a long-range amplitude correction.  Details can be found in~\cite{DAHLSTROM2013}. 
It can be verified that $\phi^{cc}_{k,\kappa} \approx -\phi^{cc}_{\kappa,k}$, i.e., the absolute cc-phase for absorption and emission between two  
given energy levels is almost identical above \unit[10]{eV} photoelectron energy.
However, there is a slight difference of $\approx 0.01$~rad at 5~eV. 

Using the asymptotic approximation, the phase of the two-photon ionization amplitudes for any particular transition channel
can be expanded as~\cite{DAHLSTROM2013} 
\begin{eqnarray} 
&\mbox{arg}[{\cal M}^{(2,e)}_{A,\ell}]&\approx -\frac{\ell_{q+1}\pi}{2}\!+\!\eta_{\ell_{q+1}}\!+\! \phi_{f,q+1}^{cc}; \\
&\mbox{arg}[{\cal M}^{(2,a)}_{B,\ell}]&\approx -\frac{\ell_{q-1}\pi}{2}+\!\eta_{\ell_{q-1}}\!+\! \phi_{f,q-1}^{cc}. 
\end{eqnarray} 
The Wigner-like phase $\eta$ depends on the angular momentum ($\ell_{q+1}$ or $\ell_{q-1}$) and the energy ($\epsilon_{q+1}$ or $\epsilon_{q-1}$) of the intermediate state 
reached upon the XUV absorption, while $\phi^{cc}$ 
depends only on the energy of the two states involved in the free-free transitions. In special cases, where only one angular momentum 
channel \hbox{($\ell_{q+1}=\ell_{q-1}=\lambda$)}  is significant in the bc-transition, the atomic phase shift can be written as
\begin{align} 
\Delta\phi_{atom}=\mbox{arg}\left[\sum_{\ell}{\cal M}^{(2,e)}_{A,\ell}{\cal M}^{*(2,a)}_{B,\ell}\right]  \approx \Delta\eta_\lambda+\Delta\phi^{cc} 
\label{1sb_asym}
\end{align} 
with \hbox{$\Delta\eta_\lambda \!=\! \eta_{\lambda}(\epsilon_{q+1})\!-\!\eta_{\lambda}(\epsilon_{q-1})$} and  \hbox{$\Delta\phi^{cc} \!=\!  \phi_{f,q+1}^{cc}\!-\!\phi_{f,q-1}^{cc}$} 
denoting the Wigner and cc phase differences, respectively. 

Equation~(\ref{1sb_asym}) is the same as Eq.~(\ref{eq:1-SB_AsymApprx}) and is broadly used to measure the Wigner delay in various systems.
We emphasize, however, that $\phi^{cc}$ here does not depend on the different angular momenta involved in the cc-transitions, 
and the bc-transition step contains only one dominant angular channel. 
In cases where several angular-momentum channels are involved in the bc-transition or $\phi^{cc}$ depends on the angular momenta of the final states, 
the phase extracted from RABBITT experiments depends on the detection angle of the observed electron. 
The atomic phase $\Delta\phi_{atom}$ measured in angle-integrated \hbox{RABBITT} schemes then becomes the weighted average of the Wigner and cc-phases of each contributing channel.

\subsection{Decomposition Approximation}\label{subsec:Decomposition}
For \hbox{RABBITT} schemes involving more than two photon transitions (\hbox{$N > 2$}),
the calculation of the necessary matrix elements becomes an increasingly formidable task. To get around this difficulty, we apply the 
ideas of the asymptotic approximation to estimate the phases of the higher-order matrix elements. Details of our derivation are provided in the Appendix.
This allows us to decompose the phase of higher-order matrix elements into a sum of the phases generated by several subsequent single-photon  transitions: 
\begin{eqnarray}
\mbox{arg}[{\cal M}_{P,\ell}^{(N)}] \approx &&~\frac{(N-2)\pi}{2} -\frac{\lambda\pi}{2}+\eta_{\lambda}+\phi_{k_2,k_1}^{cc}+\phi_{k_3,k_2}^{cc}\nonumber\\
&&~+ ... + \phi_{k_{N-2},k_{N-1}}^{cc}+\phi_{k_{N},k_{N-1}}^{cc}.
\label{eq:genFakApprox}
\end{eqnarray}

The decomposition approximation can be interpreted as a step-wise build-up of the final phase, starting with promotion from the initial bound state to a continuum state by the XUV and
followed by $N-1$ transitions within the continuum states driven by the NIR. We emphasize that this approximation requires that we only have 
one bc-channel ($\lambda$) and $\phi^{cc}$ does not depend on $\ell$.

\subsection{3-SB \hbox{RABBITT}}\label{subsec:3SB}
As an example, we now apply the decomposition approximation to the \hbox{3-SB} \hbox{RABBITT} case.
In this scheme, the consecutive harmonic peaks in the photo\-electron spectrum are separated by four times the probe photon energy ($\omega$). 
Figure~\ref{fig:trans}~(b) shows the \hbox{3-SB} \hbox{RABBITT} scheme and the dominant transition paths, up to fourth order, that are involved in the formation of three sidebands.
The population of the center sideband $S_c$ requires the absorption of an XUV pump photon ($\Omega$) and the exchange of 
at least two probe photons. There are two dominant paths (H and~I) leading to $S_c$, which can both be described using 3rd-order matrix elements. 

On the other hand, to explain the oscillations of the lower sideband $S_l$ and the higher sideband $S_h$, one needs to consider 
4th-order dipole transitions, since at least one such high-order process has to be involved 
(e.g., path~C for $S_l$ and path N for $S_h$). 
Altogether,  there are five transition terms involved: one 2nd-order and four 4th-order terms.
Interference among the paths D to G, or similarly J to M, however, does not result in delay-dependent oscillations. 
Furthermore, the relation \hbox{$\phi_{k,\kappa}^{cc}=-\phi_{\kappa,k}^{cc}$}, along with the decomposition approximation, 
implies that the phases from back and forth transitions between the same two energy levels will cancel out. 
Hence, apart from a trivial additional $\pi$ shift, 
the phases in all fourth-order absorption paths (E, F, and G) 
would be the same as in the 2nd-order absorption path~(D). 
Similarly, there would be no phase difference between the paths J, K, L, and M.  
Within this approximation, therefore, we can ignore the higher-order paths E to G and K to M, as they will 
only change the amplitude but not the phase of the oscillation.
This results in the following equations:
\begin{align}
S_l(\!\tau,k_l\!)\! &\propto\!\! \sum_\ell \left|\tilde{E}_{q+1}\tilde{E}^{*3}_{\omega} {\cal M}^{(4,e)}_{C,\ell}(k_l) + \tilde{E}_{q-1}\tilde{E}_{\omega} {\cal M}^{(2,a)}_{D,\ell}(k_l) \right|^{2} \nonumber\\
&= I^l_0 + I^l_1 \, \cos(4\,\omega\tau -\Delta\phi_{XUV}-\Delta\phi^l_{atom});\\
S_c(\!\tau,k_c\!)\! &\propto\!\! \sum_\ell \left|\tilde{E}_{q+1}\tilde{E}^{*2}_{\omega} {\cal M}^{(3,e)}_{H,\ell}(k_c) + \tilde{E}_{q-1}\tilde{E}^2_{\omega} {\cal M}^{(3,a)}_{I,\ell}(k_c) \right|^{2} \nonumber\\
& = I^c_0 + I^c_1 \, \cos(4\,\omega\tau -\Delta\phi_{XUV}-\Delta\phi^c_{atom});\\
S_h(\!\tau,k_h\!)\! &\propto\!\! \sum_\ell \left|\tilde{E}_{q+1}\tilde{E}^{*}_{\omega} {\cal M}^{(2,e)}_{J,\ell}(k_h) + \tilde{E}_{q-1}\tilde{E}^3_{\omega} {\cal M}^{(4,a)}_{N,\ell}(k_h) \right|^{2} \nonumber\\
& = I^l_0 + I^l_1 \, \cos(4\,\omega\tau -\Delta\phi_{XUV}-\Delta\phi^h_{atom}).
\end{align}
By applying the decomposition approximation to the various atomic phase 
contributions in the above equations, and again assuming that there is only one partial wave ($\lambda$) created in the XUV ionization process, 
these phases can be written as
\begin{eqnarray}
\Delta\phi^l_{atom}&=&\mbox{arg}\left[\sum_{\ell}{\cal M}_{C,\ell}^{(4,e)}\!{\cal M}_{D,\ell}^{*(2,a)}\right]\!\nonumber\\ 
&\approx& \!\Delta\eta_\lambda+\phi_{h,q+1}^{cc}\!+\phi_{c,h}^{cc}+\!\phi_{l,c}^{cc}-\phi_{l,q-1}^{cc}+\pi; \label{eq:3SBfact_a}\\
\Delta\phi^c_{atom}&=&\mbox{arg}\left[\sum_{\ell}{\cal M}_{H,\ell}^{(3,e)}\!{\cal M}_{I,\ell}^{*(3,a)}\right]\nonumber\\
& \!\approx& \!\Delta\eta_\lambda+\phi_{h,q+1}^{cc}+\phi_{c,h}^{cc}-\phi_{c,l}^{cc}-\phi_{l,q-1}^{cc};  \label{eq:3SBfact_b}\\
\Delta\phi^h_{atom}&=&\mbox{arg}\left[\sum_{\ell}{\cal M}_{J,\ell}^{(2,e)}\!{\cal M}_{N,\ell}^{*(4,a)}\right]\nonumber\\
& \!\approx& \!\Delta\eta_\lambda+\phi_{h,q+1}^{cc}-\phi_{h,c}^{cc}-\phi_{c,l}^{cc}-\phi_{l,q-1}^{cc}-\pi. \label{eq:3SBfact_c}
\end{eqnarray}
Inserting the relation $\phi_{k,\kappa}^{cc} = -\phi_{\kappa,k}^{cc} $ into Eqs.~(\ref{eq:3SBfact_a}), (\ref{eq:3SBfact_b}), and (\ref{eq:3SBfact_c}),
we see that the atomic phases in all three sidebands are the same, except for an additional phase of~$\pi$ due to the fact that the higher and lower
sidebands are created, respectively, by interference of two- and four-photon transitions, while the center sideband is created by two three-photon
transitions.  Note, however, that the final kinetic energies of the photo\-electrons in the three sidebands are different. 

\begin{figure*}[htp]
	\centering 
	\includegraphics[width=18cm,keepaspectratio]{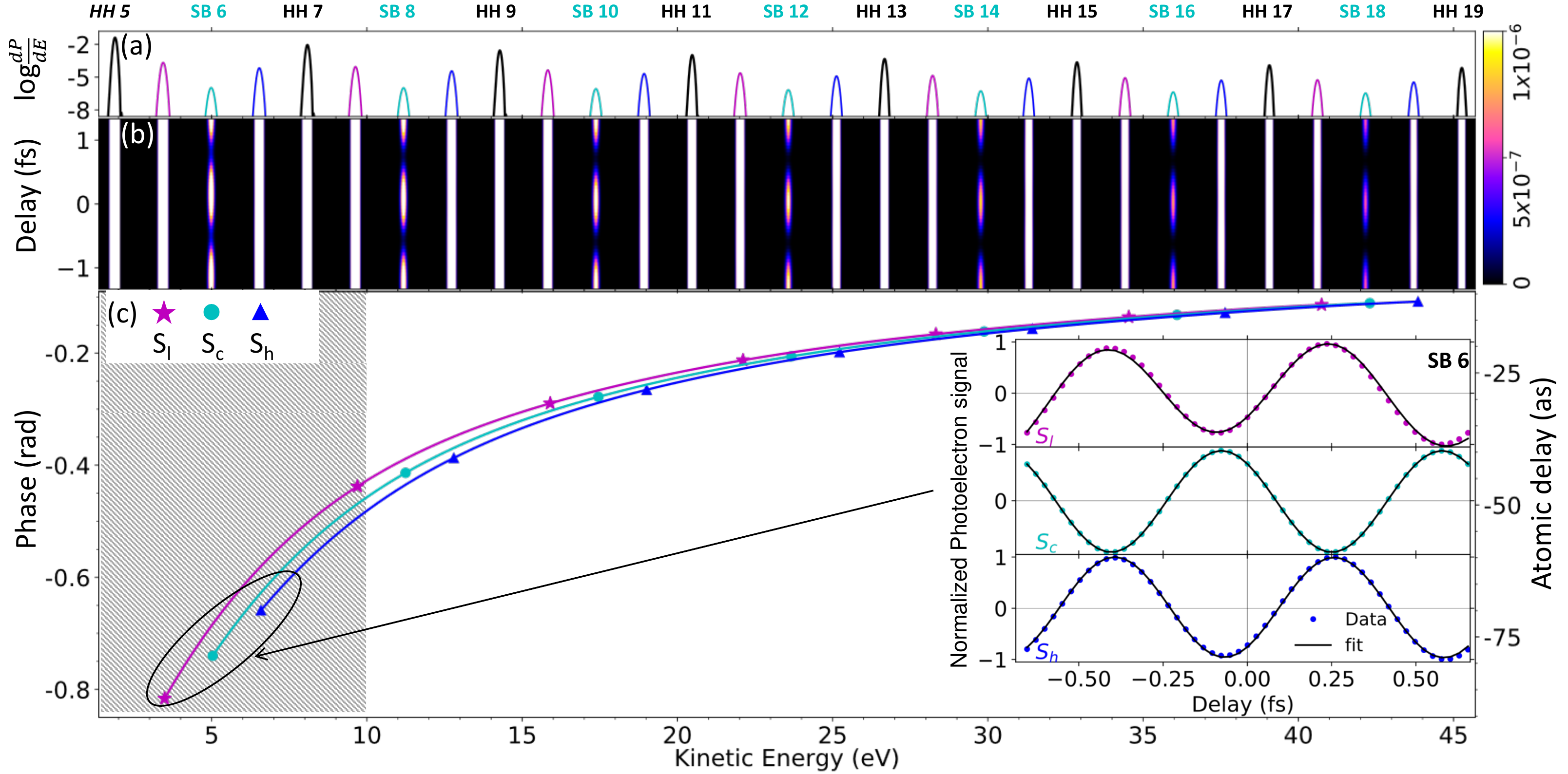}
	\caption{(a)~Photo\-electron spectrum at maximum overlap between pump and probe ($\tau=0$). \hbox{(b)~Contour} plot of a \hbox{3-SB} \hbox{RABBITT} spectrum. 
	(c)~Retrieved phases from the sideband oscillations. Inset: Fit of a cosine function (black line) 
	to the photo\-electron signals (dots) shown for the 6th sideband group order. After retrieving the sideband phases (procedure demonstrated in the inset), 
	the additional $\pi$ shift was removed from $S_l$ and $S_h$ in panel~(c) for clarity of presentation. $\Delta\phi_{XUV}=0$ in this case.
	The conversion from phase (left axis) to time (right axis) is ${\rm time = phase} / (4\,\omega)$.}
	\label{fig:3-SBR}
\end{figure*}

\section{Numerical Calculations}\label{sec:Numerics}
In order to test the validity of the decomposition approximation, we performed {\it ab initio} calculations on 
atomic hydrogen to examine the delay-dependent variation in the sideband signals for the \hbox{3-SB} \hbox{RABBITT} scheme. 
We chose a pump pulse containing eight odd harmonics ($5, 7, 9, \ldots, 19$) of the generating \unit[400]{nm} pulse.
The pulse duration of each single harmonic ($E_{q-1}$) is \unit[20]{fs} and the peak intensity is $\unit[10^9]{W/cm^2}$. 
All harmonics are in phase, i.e., \hbox{$\Delta\phi_{XUV}=0$}. The center wavelength of the probe pulse is \unit[800]{nm}, 
the pulse duration is also \unit[20]{fs}, and the peak intensity is $\unit[10^{11}]{W/cm^2}$. 

The calculations were performed with a further improved version of the computer code described by Douguet {\it et al.}~\cite{PhysRevA.93.033402}.
We performed extensive checks to ensure convergence of the predictions with the number of partial waves included, 
independence of the results from both the radial and the time steps in the discretization, and excellent agreement
between the results obtained in either the length or the velocity form of the electric dipole operator.
Given these rigorous tests, we are confident that the numerical predictions are 
highly accurate for this non\-relativistic one-electron problem and hence can be used to draw reliable conclusions
about the validity (or the lack thereof) of the approximations outlined above.  

\section{Results and Discussion}\label{sec:Results}
Figure~\ref{fig:3-SBR}~(a) exhibits the photo\-electron spectrum at zero time delay $(\tau=0)$ on a logarithmic scale. 
The pump-probe delay-resolved \hbox{RABBITT} scan is shown in Fig.~\ref{fig:3-SBR}~(b).
Since the color bar was set to make the intensity oscillations in the center sideband ($S_c$) visible, 
the other sideband oscillations are saturated with this setting.
The inset in Fig.~\ref{fig:3-SBR}~(c) shows the normalized data and the fits of the oscillatory part for the three sidebands in the lowest SB group.
Note that the phase of the center sideband is $\pi$ out of phase as shown by Eq.~(\ref{eq:3SBfact_b}).
All the phases were retrieved from the dataset shown in Fig.~\ref{fig:3-SBR}~(b). 
For the data analysis, the sideband signals were integrated over an energy window of \unit[0.25]{eV} around the peaks. 
A constant component was subtracted from the integrated sideband signals and then renormalized.  
To retrieve the phase information, the data were then fit to a function containing a cosine term and a 
quadratic term to account for the decay of the signal with the delay (see the inset in Fig.~\ref{fig:3-SBR}).
To simplify the comparison, the additional~$\pi$ phases in $S_l$ and $S_h$ were removed.

As expected, the contrast of the oscillation is best in the center sideband, because both pathways, H and I, 
contribute to its population at the same (3rd) order.  
Since the modulations in the lower and higher sidebands originate from the interference of 2nd-order and 
4th-order transition terms, the depth of the corresponding oscillations is shallow before renormalization. 
In Fig.~\ref{fig:3-SBR}~(c), 
the phases of the sidebands at their respective kinetic energies are plotted. 

\begin{figure}[b] 
	\includegraphics[width=8cm,keepaspectratio]{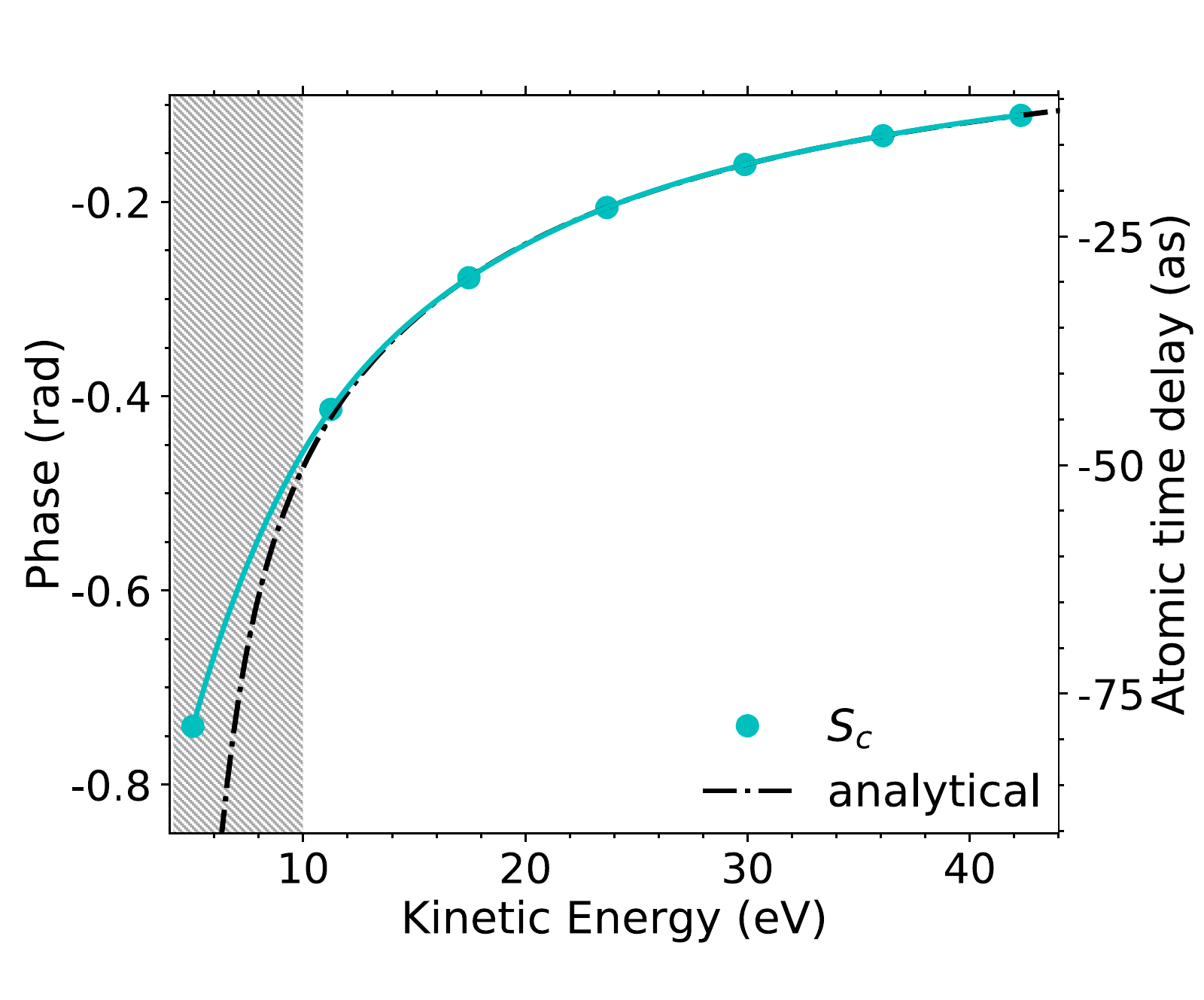}
	\caption{Phase of the $S_c$ sideband (solid circles) extracted from the TDSE calculation (see Fig.~\ref{fig:3-SBR}) 
	and the corresponding analytical phase (dot-dashed line) estimated from Eq.~(\ref{eq:3SBfact_b}). The difference above \unit[12]{eV} 
	between the two curves corresponds to a time delay smaller than \unit[0.01]{as}.}
	\label{fig:ana}
\end{figure}

Figure \ref{fig:ana} shows the phase of the center sideband and the corresponding analytical phase, as expressed in Eq.~(\ref{eq:3SBfact_b}).
The single photon ionization phase $\eta_\lambda$ for a hydrogenic system is simply the Coulomb phase:
\begin{equation}
\eta_\lambda(\kappa,Z)= \mbox{arg}\left[ \Gamma\left( \lambda+1 - {i}\frac{Z}{\kappa} \right) \right]. 
\end{equation}
Here, $\kappa$ is the momentum of the released photo\-electron, while  $\lambda$ is the orbital angular-momentum quantum number.
In the case of atomic hydrogen, there is only one transition channel available by XUV ionization, which is $s \rightarrow p$, i.e., $\lambda=1$.
The single-photon cc-phase contributions $\phi^{cc}$ for each transition are calculated using Eq.~(\ref{eq:phi_cc}). 
For more details, see Ref.~\citep{DAHLSTROM2013}. 
By inserting the expressions for $\eta$ and $\phi^{cc}$ into Eq.~(\ref{eq:3SBfact_b}), 
the analytical phase associated with the center sideband $S_c$ is obtained. 

The agreement between the two curves is remarkable for photo\-electrons with a kinetic energy above \unit[10]{eV}. 
Near threshold, however, the curves diverge.
It should be mentioned that the analytical formula for $\phi^{cc}$ breaks down at low kinetic energies. 
Nevertheless, the good agreement between the analytical phases and the phase retrieved from the numerical calculation indicates that 
the decomposition approximation works very well for the {\it center sideband} in a \hbox{3-SB} \hbox{RABBITT} scheme. 

\begin{figure}
	\includegraphics[width=7.5cm,keepaspectratio]{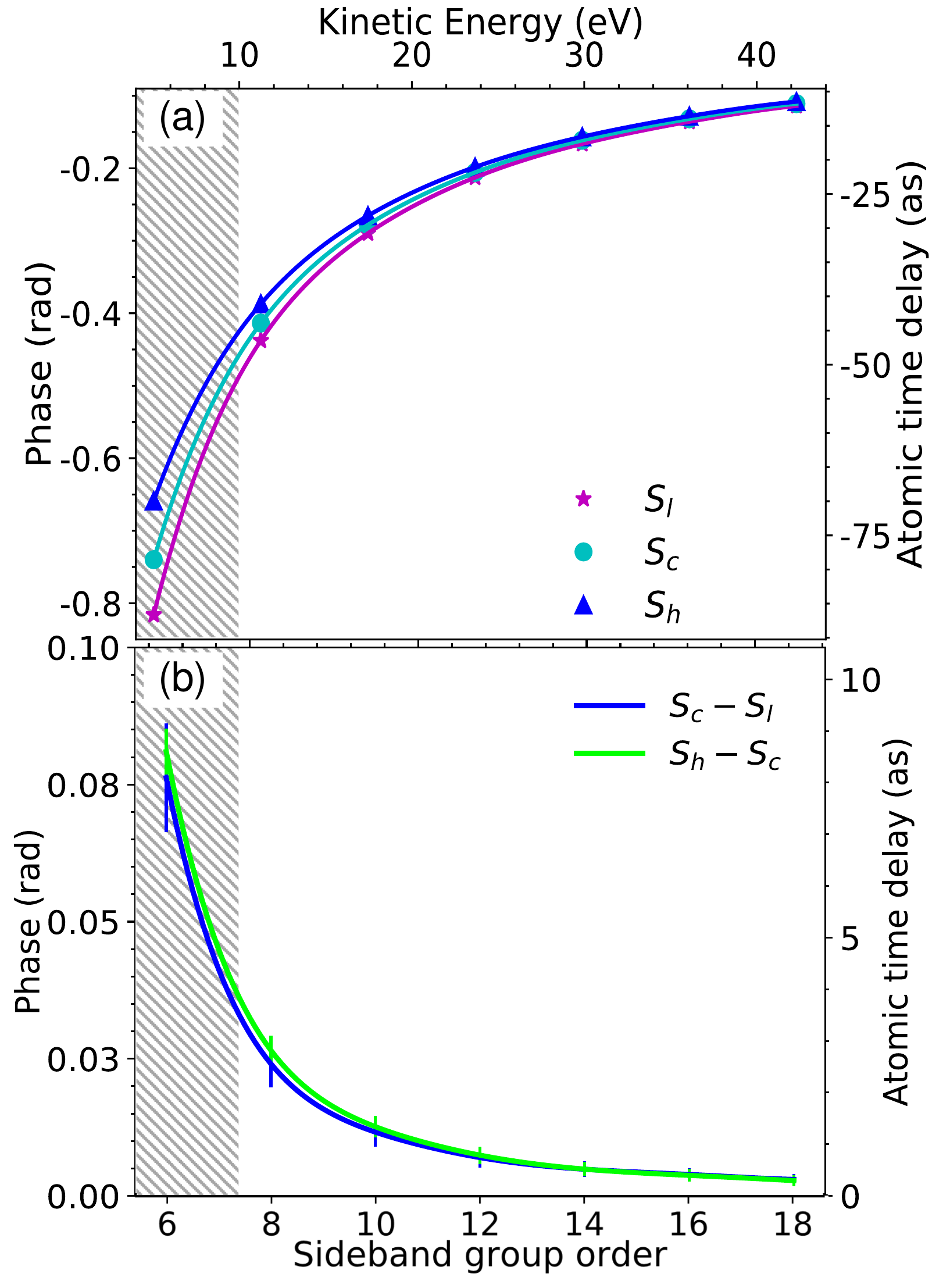}
	\caption{(a) Phases of three sidebands, $S_c$ (dots), $S_l$ (stars) and $S_h$ (triangles) over the sideband group order. 
	(b) Difference in the phases of nearby sidebands with respect to the center sideband. The error bar corresponds to the fitting error. 
	The kinetic energy on the top horizontal axis corresponds to the kinetic energy of the center sideband.}
	\label{fig:3-SBs}
\end{figure}

Figure~\ref{fig:3-SBR}~(c) is plotted again in Fig.~\ref{fig:3-SBs} (a), but now over the same sideband group order.
We immediately see that the phases in the neighboring sidebands are not identical, 
but the difference between the phases obtained from the three sidebands steadily decreases with increasing kinetic energy. 
The phase difference between  $S_h$ and $S_l$ in the same sideband group corresponds to a time delay of less than \unit[6]{as} slightly above a photoelectron energy of \unit[10]{eV}, 
but reduces to less than \unit[2]{as} beyond \unit[20]{eV}.

The fact that the phases in the three sidebands (Fig.~\ref{fig:3-SBs}) are not exactly the same could indicate that the decomposition approximation is not valid. 
However, we know from Fig.~\ref{fig:ana} that the approximation works well for the center sideband.
Therefore, we now discuss the possible origins of the discrepancy, which is clearly visible for the $S_h$ and $S_l$ sidebands.

As mentioned in the introduction, it has been shown that the actual  $\phi^{cc}$ depends on the angular momentum of the states involved in the 
transitions~\cite{Fuchs:s,ivanov2016angular,DAHLSTROM2013}. 
In that case, back and forth transitions between 
two energy levels in the continuum do not cancel out the cc-phases, i.e., $\phi_{k,\kappa}^{cc}(\ell_k) \ne -\phi_{\kappa,k}^{cc}(\ell_{\kappa})$ because 
there might be different channels involved in the two processes. 
As a consequence, the second-order absorption path (D) and the previously neglected fourth-order 
paths (E, F, and G) may not have the same phase.
Similarly, the dependence of $\phi^{cc}$ on the angular momentum may result in a 
phase difference between the paths J, K, L, and M even when the decomposition approximation holds. 

The angular-momentum dependence of $\phi^{cc}$ may also manifest itself when one considers 
only the dominant paths contributing to the modulation of all sidebands. 
If an electron with $\ell=0$ starts in the ground state, the final wave function 
of the center sideband photo\-electron after interaction with three photons will contain orbital angular quantum numbers $\ell=1$ and $\ell=3$. 
On the other hand, the lower (higher) sideband includes quantum numbers $\ell=0, 2, 4$ in the emission (absorption) path, and 
$\ell=0, 2$ in the absorption (emission) path. 
Keeping this in mind, all the $\phi^{cc}$ corresponding to the same energy levels could be different, and hence the phases in the inter\-mediate sidebands may not be the same. 

We conclude that the phase difference between the three sidebands comes most likely from the neglected $\ell$-dependence of $\phi^{cc}$ 
while the applied decomposition approximation is reasonable.
The good agreement between the $\ell$-independent analytical phase and the retrieved phases from the TDSE calculation for 
the $S_c$ (cf.\ Fig.~\ref{fig:3-SBs}) can be explained by considering the fact that both the emission and absorption paths 
are of the same order and contain the same set of possible partial waves ($p$ and $f$). As hinted in Ref.~\cite{ivanov2016angular,DAHLSTROM2013} 
for \hbox{1-SB} RABBITT calculations on He and H, the absolute phase difference between distinct partial waves for the absorption paths 
is the same as for the emission paths.  Hence the difference between the two paths nullifies the effect of the $\ell$-dependence. 
This is not true for the lower and higher sidebands, where the dominant interfering paths are of different orders.

 However, while the angular-momentum dependence is particularly prominent for low kinetic energies of the photoelectron, 
 it becomes increasingly negligible with growing kinetic energy~\cite{Fuchs:s}. 
 This parallels our observation of 
 improved agreement between the phases of all three sidebands in our \hbox{3-SB} \hbox{RABBITT} scheme. Consequently, 
 for sufficiently large photoelectron energies, where the angular momentum dependence of $\phi^{cc}$ is small, 
 our calculations support the decomposition approximation to interpret the measured atomic phase in multi-sideband RABBITT schemes as step-wise one-photon transitions. 

\section{Summary and Outlook} \label{sec:Summary}
We studied the formation of sidebands and their oscillations in the \hbox{3-SB} \hbox{RABBITT} scheme.
The phases retrieved from the oscillation of the three sidebands contain the phases of higher-order dipole matrix elements, 
which are difficult to interpret. 
A decomposition approximation was attempted to simplify the phase extraction of the higher-order matrix-elements as 
the sum of the phases of sequential one-photon transitions. 
The decomposition approximation along with the assumption that $\phi^{cc}$ is independent of the orbital angular momenta involved 
predicts that the phases extracted from all sidebands between two consecutive harmonics are the same. 

In order to check these assumptions, we performed {\it ab initio} calculations for atomic hydrogen.  
While the phases in all sideband groups are not identical, 
the difference decreases with increasing kinetic energy. 
This difference is attributed to the dependence of $\phi^{cc}$ on the angular momentum and the involvement  
of different $\ell$ channels in the three sidebands. 
We, therefore, conclude that while 
the decomposition approximation is an appropriate assumption when describing a multi-sideband \hbox{RABBITT} scheme, 
the dependence of $\phi^{cc}$ 
on the orbital angular momenta cannot be neglected for low-energy sidebands.

Multi-sideband RABBITT provides an opportunity to probe deeper into the continuum of the ionic species.
The present benchmark studies are important for a planned experiment using an Argon target, which is experimentally much more suitable than atomic hydrogen. 
We already started numerical calculations for 
this system and presented preliminary results at a recent conference~\cite{DAMOP-Ar}. 
Note, however, that having the XUV APT ionize the electron from the 3p bound orbital leads to continuum $s$ and~$d$ waves,
thereby further complicating the interpretation of the RABBITT phase.  
Work is currently in progress to address these issues.

In summary, we believe that multi-SB \hbox{RABBITT} opens up a new frontier in the study of transition phases in photo\-ionization processes.
Since many more questions remain, we
hope that the present paper will stimulate further work in this area.

\section*{Acknowledgements}
This work was supported by the Deutsche Forschungs\-gemeinschaft under \hbox{DFG-SPP-1840-HA-8399/2-1}  and
by the United States National Science Foundation under
grants No.~OAC-1834740 and PHY-1803844 (DAS, GM, KRH, KB) and PHY-2012078 (ND), as well as the XSEDE supercomputer allocation No.~PHY-090031. 
The calculations were carried out on Stampede2 and Frontera at the Texas Advanced Computing Center in Austin,
Bridges at the  Pittsburgh Supercomputing Center, and Comet at the San Diego Supercomputer Center.

\section{Appendix}
\renewcommand{\theequation}{A.\arabic{equation}}
\setcounter{equation}{0}

In this Appendix, we derive Eq.~(\ref{eq:genFakApprox}) by generalizing the asymptotic approximation for the two-photon matrix element introduced by Dahlstr\"om {\it et al.}~\citep{DAHLSTROM2013}.
In order to set the stage and introduce our short-hand notation for the often lengthy expressions, we first repeat the key ideas of the above paper and then apply them to the third and higher-order 
matrix elements.

We begin by explicitly writing Eq.~(\ref{eq:G+}) as 
\begin{widetext}
\begin{equation}
M^{(N)}(\vec{k}_N; \Omega,\omega)= -{i} \tilde{E}_{\Omega}\tilde{E}^{N-1}_{\omega}  \lim_{\varepsilon \to 0^+} 
      \int \!\! d^3\vec{k}'_{N-1} \ldots \int \!\! d^3\vec{k}'_2 \; \int \!\! d^3\vec{k}'_1  \;  
      \frac{\bra{\vec{k}_N}{z}\ket{\vec{k}'_{N-1}}...\bra{\vec{k}'_2}{z}\ket{\vec{k}'_1}\bra{\vec{k}'_1}{z}\ket{i}}{(\epsilon_{N-1}-\epsilon'_{N-1}+{i}\varepsilon)...(\epsilon_1-\epsilon'_1+{i}\varepsilon)} \label{eq:eq_M_k_2}
\end{equation}
\end{widetext}
Here $\epsilon_i$ is the initial state energy and $\epsilon_i+\Omega +n\omega=\epsilon_{n+1}$. As in~\citep{DAHLSTROM2013}, we neglect all contributions from bound states,
since they are expected to be small if the photon energy for the bc-transition is sufficiently high. 

The final and intermediate continuum states are decomposed into partial waves as
\begin{equation}
\varphi_{\vec{k}}(\vec{r})=(8\pi)^{3/2}\sum_{\ell,m} {i}^{\ell} \mathrm{e}^{-{i}\eta_{\ell}(k)}Y^*_{\ell,m}(\hat{k})Y_{\ell,m}(\hat{r})R_{k,\ell}(r),  \label{eq:partial_wave_k}
\end{equation}
where $\eta_{\ell}(k)$ denotes the scattering phases, $Y_{\ell,m}$ the spherical harmonics, and $R_{k,\ell}$ are the radial parts. 
Splitting the initial state into its radial and angular parts
according to
\begin{equation}
\varphi_{n_i,\ell_i,m_i}(\vec{r})=Y_{\ell_i,m_i}(\hat{r})R_{n_i,\ell_i}(r),  \label{eq:initial_state}
\end{equation}
using $z= \sqrt{\frac{4\pi}{3}} r Y_{1,0}(\hat{r})$, and inserting (\ref{eq:partial_wave_k}) and (\ref{eq:initial_state}) into (\ref{eq:eq_M_k_2}) yields for $N=2$
\begin{eqnarray}
M^{(2)}(\vec{k}_2; \Omega,\omega)&=&  -{i} \frac{4\pi}{3}(8\pi)^{3/2}\tilde{E}_{\Omega}\tilde{E}_{\omega}  \nonumber \\
    && \hspace{-10truemm}\times \sum_{\ell_2,m_2} (-{i})^{\ell_2} \mathrm{e}^{{i}\eta_2} Y_{\ell_2,m_2}(\hat{k}_2) \nonumber \\ 
    && \hspace{-10truemm}\times \sum_{\ell_1,m_1}\bra{Y_{\ell_2,m_2}}Y_{1,0}\ket{Y_{\ell_1,m_1}}\bra{Y_{\ell_1,m_1}}Y_{1,0}\ket{Y_{\ell_i,m_i}} \nonumber \\ 
    && \hspace{-10truemm}\times \; T^{(2)}_{\ell_2,\ell_1,\ell_i}(k_2; \Omega,\omega) \nonumber \\
    && \hspace{-10truemm} = \tilde{E}_{\Omega}\tilde{E}^2_{\omega} \sum_{\ell_2} {\cal M}^{(2)}_{\ell_2,m_2}(k_2) Y_{\ell_2}(\hat{k}_2).  \label{eq:M_l_2}
\end{eqnarray}
This defines ${\cal M}^{(2)}_{\ell_2,m_2}(k_2)$ and simplifies to Eq.~(\ref{eq:amplitude_2}) of the main text for $\ell_i = m_i = 0$ and linearly
polarized light ($m_2 = 0$).

Defining the first-order radial matrix element
\begin{equation}
T^{(1)}_{\ell_1,\ell_i}(k'_1; \Omega) = \bra{R_{k'_1,\ell_1}}r\ket{R_{n_i,\ell_i}}  
\end{equation}
allows us to write the second-order element as
\begin{eqnarray}
T^{(2)}_{\ell_2,\ell_1,\ell_i}(k_2; \Omega,\omega) &=& ~~~~~~~~~~~~~~~~~~~~~~~~~~~~~~~~~~~~~~~~~~~~~\nonumber \\
   && \hspace{-25truemm} \lim_{\varepsilon \to 0^+} \int_{0}^{\infty} d\epsilon'_1 \; \frac{\displaystyle \bra{R_{k_2,\ell_2}}r\ket{R_{k'_1,\ell_1}}T^{(1)}_{\ell_1,\ell_i}(k'_1; \Omega)}{\displaystyle \epsilon_i+\Omega-\epsilon'_1+{i}\varepsilon}.
\end{eqnarray}
The first-order perturbed wave function~\citep{DAHLSTROM2013} is defined as
\begin{eqnarray}
\ket{\rho_{k_1,\ell_1}}  &  = & {\displaystyle\lim_{\varepsilon \to 0^+}} {\displaystyle \int_{0}^{\infty}} d\epsilon'_1  \; 
    \frac{\displaystyle T^{(1)}_{\ell_1,\ell_i}(k'_1; \Omega)\ket{R_{k'_1,\ell_1}}}{\displaystyle\epsilon_1-\epsilon'_1+{i}\varepsilon}  \nonumber \\  
& = & \hbox{$\cal P$}  {\displaystyle\int_{0}^{\infty}} d\epsilon'_1 \; \frac{\displaystyle T^{(1)}_{\ell_1,\ell_i}(k'_1; \Omega)\ket{R_{k'_1,\ell_1}}}{\displaystyle \epsilon_1-\epsilon'_1} \nonumber \\  
& - &{i}\pi T^{(1)}_{\ell_1,\ell_i}(k_1; \Omega)\ \ket{R_{k_1,\ell_1}}, \label{eq:rho_def_p.v.}
\end{eqnarray}
where $\cal P$ denotes the principal value.

The key in deriving an analytic expression for the contribution to the phase of the matrix element is 
to replace the radial functions of the intermediate and final continuum states by their asymptotic form
\begin{equation}
\lim_{r \to \infty} R_{k,\ell}(r)\approx \frac{N_k}{r} \sin(kr+\phi_{k,\ell}(r))  \label{eq:sin_asm}
\end{equation}
Here $N_k\approx \sqrt{2/\pi k}(1-Z/(2rk^2))$ is the amplitude of the asymptotic wave for a long-range potential \hbox{($-Z/r$)}, and the asymptotic phase is  $\phi_{k,\ell}(r)=(Z/k)\,\mathrm{ln}(2kr)+\eta_{\ell}(k)-\pi \ell/2$.
Using the same steps and approximations as outlined in~\citep{DAHLSTROM2013}, this leads to the
approximate form 
\begin{equation}
\lim_{r \to \infty} \rho_1(r)\approx -\frac{\pi N_1}{r}\ T^{(1)}(k_1)\ \mathrm{e}^{{i}(k_1 r+\phi_1(r))}  \label{eq:rho_r_1}
\end{equation}
for the perturbed wave function after further compressing the notation by defining
$R_{k_n,\ell_n}\equiv R_n$, $\ket{\rho_{k_1,\ell_1}}\equiv \ket{\rho_1}$, \hbox{$ N_{k_n}\equiv N_n$}, $\phi_{k_n,\ell_n}(r)\equiv\phi_n(r)$, 
$ T^{(1)}_{\ell_1,\ell_i}(k_1; \Omega)\equiv T^{(1)}(k_1)$, and $\eta_{\ell_n}(k_n)\equiv \eta_n  $.

The 2nd-order matrix element then becomes
\begin{eqnarray}
T^{(2)}_2(k_2) &=& \bra{R_2}r\ket{\rho_1} \nonumber \\
               &\approx& \frac{ T^{(1)}(k_1) }{{i}\sqrt{ k_1 k_2} } \int_0^{\infty} dr\ \left(r-\frac{1}{2}\left(\frac{1}{k_1^2}+\frac{1}{k_2^2}\right)\right)  \nonumber \\
&\times& \ \sin(k_2r+\phi_2(r))  \mathrm{e}^{{i}(k_1 r + \phi_1(r))}.
\end{eqnarray}
After writing the sin-term in exponential form, dropping the fast-oscillating term containing $k_1+k_2$ while keeping the term with $k_1-k_2$, and
introducing the substitution variable $-i\,(k_1-k_2)\,r$, one obtains a $\Gamma$-function with complex argument.
Using this function, we find
\begin{eqnarray}
T^{(2)}_2(k_2) &\approx& \frac{ T^{(1)}(k_1) }{\sqrt{ k_1 k_2} } \; \frac{\mathrm{e}^{-Z(1/k_1-1/k_2)\pi/2}}{(k_1-k_2)^2} \ {i}^{\ell_2-\ell_1+1}\mathrm{e}^{{i}(\eta_1-\eta_2)}  \nonumber \\
     &\times&   \frac{(2k_1)^{{i}Z/k_1}}{(2k_2)^{{i}Z/k_2}} \frac{(\Gamma[{2+{i}Z(1/k_1-1/k_2)}]+\gamma(k_2,k_1))}{(k_1-k_2)^{{i}Z(1/k_1-1/k_2)}} \label{eq:T_2_final} \nonumber \\
\end{eqnarray}
with $\gamma(k_2,k_1)=\frac{{i}Z(k_1-k_2)}{2}\left(\frac{1}{k_1^2}+\frac{1}{k_2^2}\right) \Gamma[{1+{i}Z(1/k_1-1/k_2)}]$ 
accounting for the effect of the long-range potential~\citep{DAHLSTROM2013}.
The phase of the radial matrix element is 
\begin{equation}
\mathrm{arg}[T^{(2)}_{\ell_2,\ell_1,\ell_i}(k_2)] = \frac{\pi}{2}(\ell_2-\ell_1+1)+(\eta_1 -\eta_2 ) +\phi^{cc}_{2,1},\label{eq:Phase_T2}
\end{equation}
where
\begin{eqnarray}
\phi^{cc}_{2,1}&\equiv&\phi^{cc}_{k_2,k_1} \nonumber\\
 && \hspace{-10truemm}= \mathrm{arg}\left[ \frac{(2k_1)^{{i}Z/k_1}}{(2k_2)^{{i}Z/k_2}} \frac{(\Gamma[{2+{i}Z(1/k_1-1/k_2)}]+\gamma(k_2,k_1))}{(k_1-k_2)^{{i}Z(1/k_1-1/k_2)}} \right],\nonumber\\
\end{eqnarray}
and we have used that $T^{(1)}_{\ell_1,\ell_i}(k_1)$ is real.
Substituting (\ref{eq:T_2_final}) back into (\ref{eq:M_l_2}), the phase of the matrix element for the transition path $\ell_i \to \ell_1 \to \ell_2$ is
\begin{eqnarray}
\mathrm{arg}[{\cal M}^{(2)}_{\ell_2,\ell_1,\ell_i}(k_2)] & \approx & \frac{-\pi}{2}-\frac{\pi \ell_2}{2}+\eta_2 + \frac{\pi}{2}(\ell_2-\ell_1 +1) \nonumber \\
    & + & \eta_1 -\eta_2 +\phi^{cc}_{2,1} +T^{(1)}_{\ell_1,\ell_i}(k_1) \nonumber \\
    & = & - \frac{\pi\ell_1}{2}+\eta_1+\phi^{cc}_{2,1}.
\end{eqnarray}
Note the cancellations in this formula, particularly the contributions from both $i^{\ell_2}$ and $\eta_2$.  
As will be seen below, these cancellations are a general pattern when we move to higher-order matrix elements. 

Our approximation for the higher-order matrix elements is based on the above formalism.
Starting with the 3rd-order element, the equivalent of~(\ref{eq:M_l_2}) is
\begin{eqnarray}
M^{(3)}(\vec{k_3}; \Omega,\omega) & = &  -{i} \left(\frac{4\pi}{3}\right)^{3/2}(8\pi)^{3/2}\tilde{E}_{\Omega} \tilde{E}^2_{\omega}  ~~~~~~~~~\nonumber \\
&& \hspace{-20truemm} \times \sum_{\ell_3} (-{i})^{\ell_3} \mathrm{e}^{{i}\eta_3} Y_{\ell_3}(\hat{k}_3) \nonumber \\ 
&& \hspace{-20truemm} \times   \sum_{\ell_2,\ell_1}\bra{Y_{\ell_3}}Y_{1,0}\ket{Y_{\ell_2}}\bra{Y_{\ell_2}}Y_{1,0}\ket{Y_{\ell_1}}\bra{Y_{\ell_1}}Y_{1,0}\ket{Y_{\ell_i}} \nonumber \\
&& \hspace{-20truemm} \times \; T^{(3)}_{\ell_3,\ell_2,\ell_1,\ell_i}(k_3; \Omega,\omega) \nonumber \\
&& \hspace{-20truemm}  = \tilde{E}_{\Omega}\tilde{E}^2_{\omega} \sum_{\ell_3} {\cal M}^{(3)}_{\ell_3}(k_3) Y_{\ell_3}(\hat{k}_3). 
\end{eqnarray}
Here 
\begin{equation}
T^{(3)}_{\ell_3,\ell_2,\ell_1,\ell_i}(k_3; \Omega,\omega) =  \bra{R_3}r\ket{\rho_2}  \label{eq:T3_rho} 
\end{equation}
with the 2nd-order perturbed wavefunction 
\begin{equation}
\ket{\rho_2} = \int_{0}^{\infty} d\epsilon'_2 \frac{\ket{R'_2}\bra{R'_2}r\ket{\rho_1}}{(\epsilon_2-\epsilon'_2+{i}\varepsilon)}.
\end{equation}

In the asymptotic approximation, we obtain
\begin{equation}
\rho_2(r)\approx -\frac{\pi N_2}{r}\ T^{(2)}(k_2)\ \mathrm{e}^{{i}(k_2 r+\phi_2(r))} \label{eq:rho_r_2}
\end{equation}
and, consequently,
\begin{eqnarray}
T^{(3)}_3(k_3)&\approx&  \frac{ T^{(2)}(k_2) }{\sqrt{ k_2 k_3} } \frac{\mathrm{e}^{-Z(1/k_2-1/k_3)\pi/2}}{(k_2-k_3)^2} \ {i}^{\ell_3-\ell_2+1} \mathrm{e}^{{i}(\eta_2-\eta_3)}  \nonumber \\
&\times& \frac{(2k_2)^{{i}Z/k_2}}{(2k_3)^{{i}Z/k_3}} \frac{(\Gamma[{2+{i}Z(1/k_2-1/k_3)}]+\gamma(k_3,k_2))}{(k_2-k_3)^{{i}Z(1/k_2-1/k_3)}}. \nonumber \\
\end{eqnarray}
Looking at the phases,
\begin{eqnarray}
\mathrm{arg}[T^{(3)}_{\ell_3,\ell_2,\ell_1,\ell_i}(k_3)] \nonumber \\
  &&  \hspace{-20truemm} = \frac{\pi}{2}(\ell_3-\ell_2+1)+(\eta_2 -\eta_3 ) +\phi^{cc}_{3,2} + \mathrm{arg}[T^{(2)}(k_2)] \nonumber \\
  &&  \hspace{-20truemm} = \frac{\pi}{2}(\ell_3-\ell_1+2)+(\eta_1-\eta_3 ) +\phi^{cc}_{3,2}  +\phi^{cc}_{2,1}.
\end{eqnarray}
and, therefore, since the above radial matrix element is independent of the intermediate angular momentum $\ell_2$, we obtain that
\begin{eqnarray}
\mathrm{arg}[{\cal M}^{(3)}_{\ell_3,\ell_2,\ell_1,\ell_i}(k_3)] &=& \frac{-\pi}{2}-\frac{\pi \ell_3}{2}+\eta_3 + \mathrm{arg}[T^{(3)}(k_3)] ~~~~~~\nonumber \\
&=& \frac{\pi}{2} - \frac{\pi\ell_1}{2}+\eta_1+\phi^{cc}_{3,2} +\phi^{cc}_{2,1}. 
\end{eqnarray}
By repeating the procedure, i.e., straightforward induction, the phase of the $N$th-order matrix element can be written as
\begin{eqnarray}
\hspace{-10truemm} \mathrm{arg}[{\cal M}^{(N)}_{\ell_N;\ell_1}(k_N)] &=& \frac{(N-2)\pi}{2}- \frac{\pi\ell_1}{2}+\eta_{\ell_1} \nonumber \\
  &&  \hspace{-2truemm} + \; \phi^{cc}_{k_2,k_1} +\phi^{cc}_{k_3,k_2} ...+\phi^{cc}_{k_N,k_{N-1}}.
\end{eqnarray}
The analytical form of $\mathrm{arg}[{\cal M}^{(N)}_{\ell_N;\ell_1}(k_N)] $ only depends on the angular momenta of the first intermediate state after the XUV step.

\bibliographystyle{apsrev4-1}

\end{document}